\RequirePackage[2020-02-02]{latexrelease}
\documentclass[aps,prc,twocolumn,floatfix,superscriptaddress]{revtex4}
\usepackage{epsfig}
\usepackage{amsmath}
\usepackage{color}
\usepackage{graphicx}
\begin{document}

\title{Effect of Isolation Criteria on Prompt Photon Production in Relativistic Nuclear Collisions}

\author{\large Sinjini Chandra}
\email{s.chandra@vecc.gov.in}
\author{\large Rupa Chatterjee}
\email{rupa@vecc.gov.in}
\author{\large Zubayer Ahammed}

\affiliation{Variable Energy Cyclotron Centre, 1/AF, Bidhan Nagar, Kolkata-700064, India}
\affiliation{Homi Bhabha National Institute, Training School Complex, Anushaktinagar, Mumbai 400094, India}

\begin{abstract}
Prompt photon measurements in relativistic nuclear collisions serve as an essential comparative basis for heavy ion studies enabling the separation of medium induced effects.
However, the identification of prompt photons is experimentally challenging due to substantial backgrounds from photons produced in hadron decays and jet fragmentation. Appropriate isolation criteria are applied to suppress these background contributions.
We analyze prompt photon spectra using the JETPHOX framework to quantify the relative contributions of fragmentation and direct production mechanisms to the total photon yield. We perform a systematic study of the impact of isolation criteria on prompt photon production in relativistic nuclear collisions with emphasis on their dependence on beam energy and photon transverse momentum.
The fragmentation contribution is found to be substantially large particularly for $p_T < 15$ GeV and the isolation criterion plays a crucial role in the analysis of prompt photons in that $p_T$ region. A dynamical isolation criterion suppresses the fragmentation component more effectively than a fixed one in this region. Furthermore, the isolation criterion shows a stronger dependence on beam energy and photon $p_T$ than on system size. These observations emphasize the importance of employing carefully selected and consistent isolation criteria when comparing experimental data with theoretical calculations especially for observables sensitive to fragmentation.
\end{abstract}


\maketitle

\section{Introduction} 

Relativistic heavy ion collisions provide an unique opportunity to explore the properties of the quark-gluon plasma (QGP), a strongly interacting matter created under extreme conditions of temperature and energy density~\cite{qgp1, qgp2, qgp3, qgp4, kolb}. Among the various probes available, direct photons are known to be one of the most versatile and clean tools for studying the initial state as well as the space-time evolution of the produced system~\cite{feinberg, qgp_phot, phot1, phot2, phot3, phot4, gabor, rhic-lhc, pp_510_oscar, phot_jet, t1}. 
Prompt photons originating from initial hard scatterings in heavy ion ($AA$) collisions populate the direct photon spectrum in the high $p_T$ ($>3$ GeV) region~\cite{b0}. Whereas, thermal photons dominate the relatively lower $p_T$ region~\cite{enhancement, pramana}. Prompt photons produced in proton-proton ($pp$) collisions are appropriately normalized and compared with the direct photon spectrum in $AA$ collisions to estimate the thermal contribution from the QGP medium. Recent studies suggest that high multiplicity $p+Pb$ and $d+Au$ collisions may also create a QGP medium resulting in thermal radiation in addition to the prompt component in the direct photon spectrum~\cite{ppb}.

As prompt photons are produced prior to QGP formation, they serve as an important probe of the initial state of the collision and the underlying parton distribution functions (PDFs) of the colliding nuclei. Comparison of the prompt production from $pp$ collisions with those from $AA$ collisions enables the study of nuclear modifications to PDFs (nPDFs) including shadowing, anti-shadowing, EMC effects, etc. across a wide range of energy scale~\cite{shadow, shadow1}. Thus, prompt photons constitute a powerful probe for constraining and advancing our quantitative understanding of the partonic structure in both free protons and nuclei.

Accurate interpretation of the direct photon spectrum and anisotropic flow requires precise estimation of various photon sources with their separation and quantification remaining a central challenge in relativistic nuclear physics~\cite{residual}. Prompt photon production in relativistic nuclear collisions consists of two components, direct and fragmentation. The direct photons mostly arising from $2 \rightarrow 2$ processes such as quark–gluon Compton scattering and quark–antiquark annihilation are azimuthally symmetric. Whereas, the fragmentation photons are influenced by the medium thickness~\cite{prompt, prompt1, review}.

Prompt photons originating from the fragmentation process are typically accompanied by hadrons and the direct component of prompt photons can be isolated by suppressing this fragmentation contribution. To achieve this, the high energy experiments commonly employ an isolation cut where, a cone of radius $R$ is defined in the $(\eta,\phi)$ plane around the photon candidate~\cite{gabor}. If the total transverse energy within the cone (excluding the photon) exceeds a predefined threshold ($p_T^{\rm cut}$), the photon is classified as originating from fragmentation and is rejected from the total contribution.

The primary experimental challenge in analyzing the photon spectrum from relativistic nuclear collisions, stems from the large background of decay photons from $\pi^0$ and $\eta$ mesons, which predominantly populate the low $p_T$ region~\cite{gabor, review}. Recent advancements in decay background subtraction techniques have significantly improved the quality of photon data at both RHIC and LHC energies~\cite{phenix_pp200, phenix_auau200, alice_pp13}. Additionally, the use of isolation criteria further helps in suppressing this background. It has been shown that the shower shape analysis method is quite effective for decay background subtraction in the high $p_T$ (above 6 GeV) region at 7 and 13 TeV LHC energy~\cite{alice_pp13, alice_pp_7TeV}. Therefore, the isolation criteria used in experiments are expected to be particularly effective for decay photons in the low $p_T$ region only.


The isolation cut applied in experimental analyses cannot uniquely distinguish decay photons from fragmentation photons and it tends to suppress both. However, in theoretical calculations the isolation criterion is implemented primarily to reduce the fragmentation component.

In this study, we perform a detailed investigation of the direct and fragmentation components of prompt photons produced in relativistic nuclear collisions using the JETPHOX framework~\cite{prompt,jp1,jp3}. We quantify the relative contributions of these components across different beam energies and collision systems. 

The main objective of our study is to investigate how the isolation cuts affect the theoretical prompt photon spectra and to explore the sensitivity of the results to different isolation criteria. This may help to assess whether the comparison between theoretical calculations and experimental data remains consistent under various isolation conditions. It is to be noted a similar analysis for isolation cuts of decay photons would be valuable, especially in the low $p_T$ region.

Earlier studies have shown that, although the prompt fragmentation contribution is relatively small, it contributes to the the azimuthal anisotropy of photons in $AA$ collisions~\cite{frag_v2} and thereby may provide valuable insights into the ‘direct photon puzzle'~\cite{cfhs, cs, puzzle, flow_phenix, direct}. Thus, understanding how experimental isolation definitions influence theoretical prompt photon yields is essential for meaningful comparisons between data and theory.

\begin{figure}
        \includegraphics[width=90mm,trim={0 0 0 0},clip]{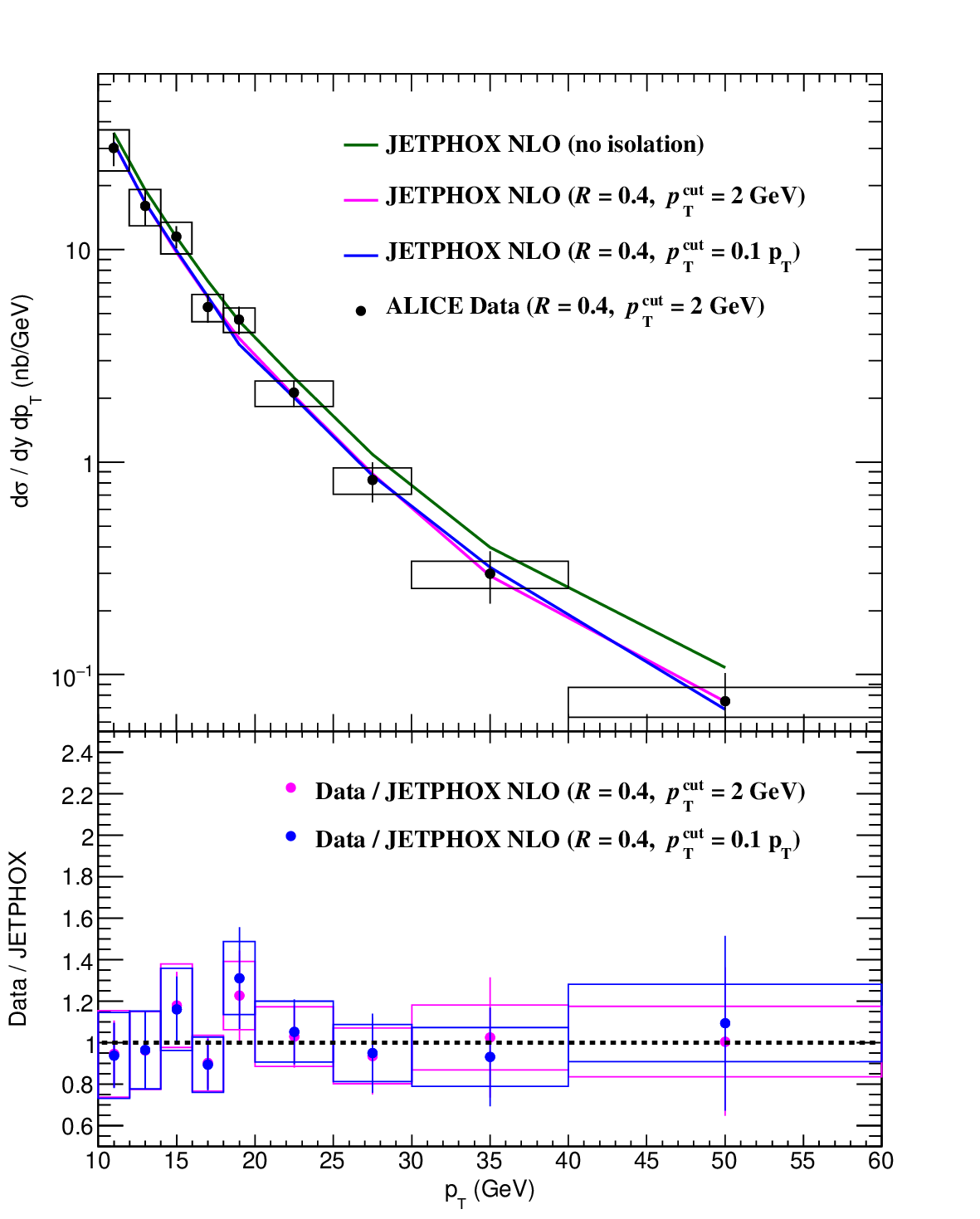}  
        \caption{\label{fig1} Prompt photon cross section in $pp$ collisions at $\sqrt{s}=7$ TeV for two different isolation criteria and for the case without any isolation cut along with ALICE data~\cite{alice_pp_7TeV} in the rapidity range $|\eta|<0.27$.}
\end{figure}

\begin{figure}
        \includegraphics[width=90mm,trim={0 0 0 0},clip]{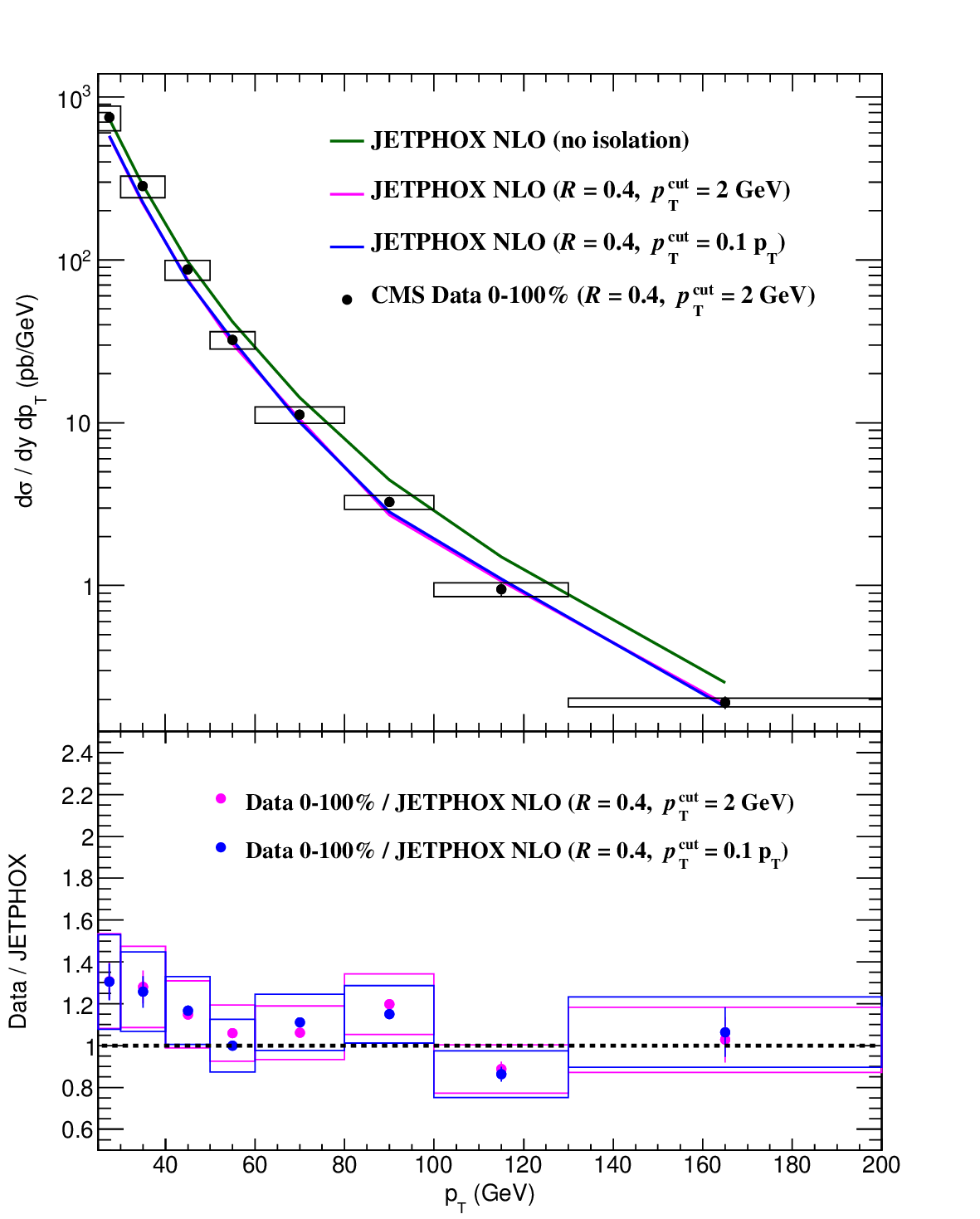} 
        \caption{\label{fig2} Prompt photon cross section in $Pb+Pb$ collisions at $\sqrt{s_{NN}}=5.02$ TeV for two isolation criteria and for the case without any isolation cut along with CMS data~\cite{cms_PbPb_5TeV} in the rapidity range $|\eta|<1.44$.}
\end{figure}

\begin{figure}
        \includegraphics[width=90mm,trim={0 0 0 0},clip]{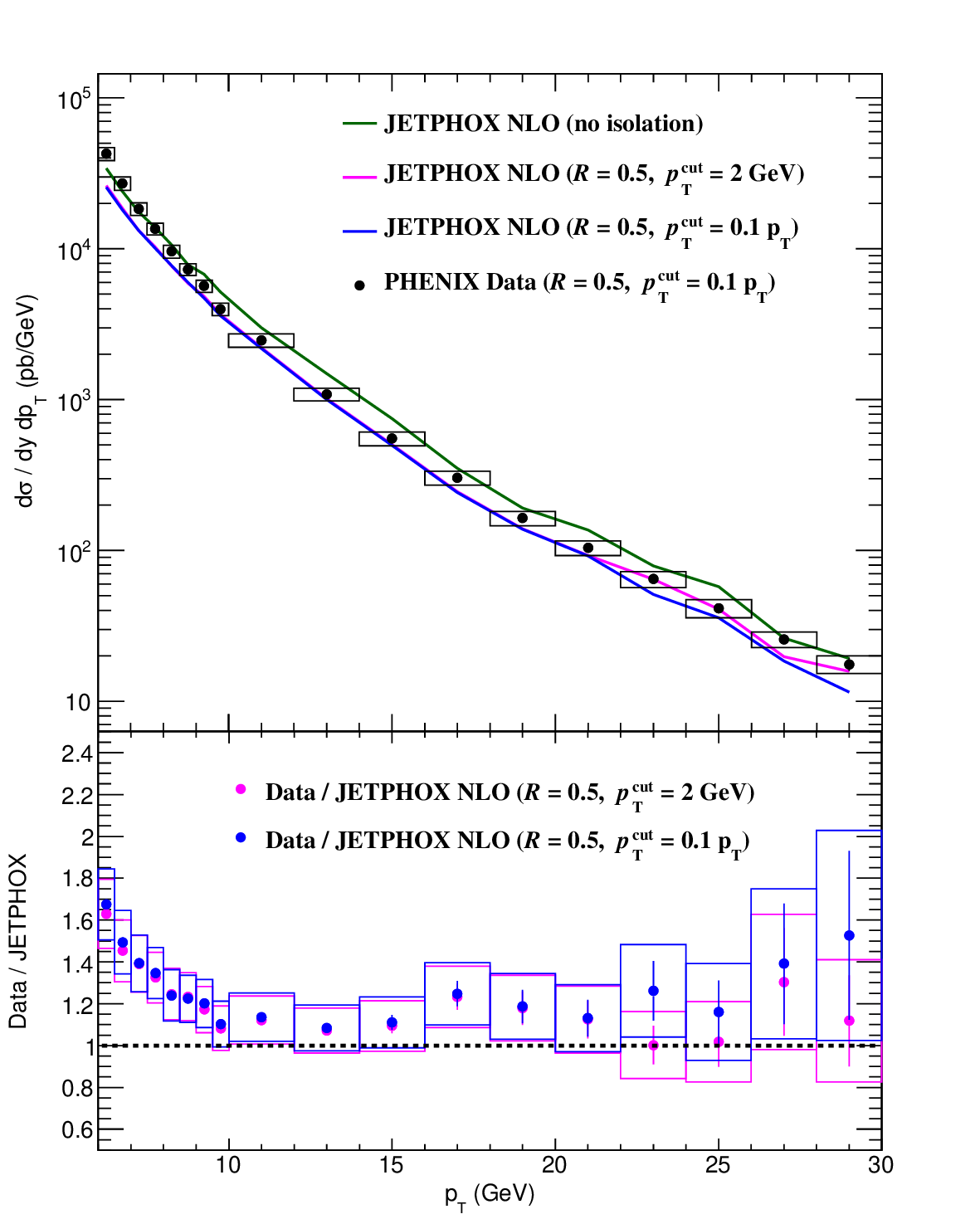}  
        \caption{\label{fig2.1} Prompt photon cross section in $pp$ collisions at $\sqrt{s}=510$ GeV for two different isolation criteria and for the case without any isolation cut along with PHENIX data~\cite{phenix_pp_510GeV} in the rapidity range $|\eta|<0.25$. The isolation cone radius $R$ is taken as 0.5, same as in the PHENIX data analysis.}
\end{figure}

\section{Prompt photon production using JETPHOX}
The production rate of prompt photons comprising both direct and fragmentation components, can be computed using next-to-leading order (NLO) perturbative QCD (pQCD) calculations.
The leading order (in $\alpha \alpha_s$) contribution to direct prompt photon production arises from the Born level processes $q\bar{q} \rightarrow \gamma g$ and $q(\bar{q}) g \rightarrow \gamma q(\bar{q})$. The NLO calculation introduces contributions from sub-processes such as $q\bar{q} \rightarrow \gamma gg$, $gq(\bar{q}) \rightarrow \gamma gq(\bar{q})$, as well as from virtual corrections to the Born level diagrams~\cite{jp1, nlo_prompt, p2}. In addition, the evaluation of these higher order corrections generates the LO fragmentation contribution where the photon originates from the collinear fragmentation of a hard parton.

The total prompt photon production cross-section can be expressed as the sum of two distinct terms~\cite{prompt}:,
\begin{equation}
\frac {d \sigma} {d \overrightarrow p_T d \eta} = {\frac {d \sigma^{(D)}} 
{d \overrightarrow p_T d \eta}} + {\frac {d \sigma^{(F)}} {d 
\overrightarrow p_T d \eta}}
\label{prompt}
\end{equation}
In the above equation, `D' denotes the direct contribution while, `F' represents the fragmentation component. The `D' photons are well separated and isolated from hadronic activity, whereas `F' photons originate from the collinear fragmentation of high $p_T$ partons and are therefore accompanied by hadrons. This distinction can be exploited by applying isolation cuts, which effectively separate direct photons from the fragmentation background.

The two terms in Eqn.~(\ref{prompt}) can be written explicitly as~\cite{prompt1}:
\begin{eqnarray}
\frac {d \sigma^{(D)}} {d \overrightarrow p_T d \eta} &=& \sum_{i,j=q, \bar{q},g} \int dx_{1} dx_{2} 
\nonumber\\
& \times & \ F_{i/h_1}(x_{1},M)\ F_{j/h_2}(x_{2},M) {\frac { \alpha_s (\mu_R)}{2 \pi}}
\nonumber\\
& \times& \left 
({\frac {d \hat{\sigma}_{ij}}{d \overrightarrow p_T d \eta}} +{\frac 
{ \alpha_s (\mu_R)} {2 \pi}} K_{ij}^{(D)}  (\mu_{R},M,M_{_F})\right )  \ 
\end{eqnarray}
and
\begin{eqnarray}
\frac {d \sigma^{(F)}} {d \overrightarrow p_T d \eta} &=& \sum_{i,j,k=q,
\bar{q},g} \int dx_{1} dx_{2} {\frac{dz}{z^2}} \ F_{i/h_1}(x_{1},M) \nonumber \\
 && 
\times \ F_{j/h_2}(x_{2},M) \ D_{\gamma/k}(z, M_{_F})  \ ({\frac { \alpha_s (\mu_R)}{2 \pi}})^2
\nonumber \\
&& \left( {\frac {d \hat\sigma_{ij}^k}{d \overrightarrow p_T d \eta}} +{\frac { \alpha_s 
(\mu_R)}{2 \pi}} K_{ij,k}^{(F)} (\mu_{R},M,M_{_F}) \right).
\end{eqnarray}
Here $F_{i/h_{1,2}}(x,M)$ are the parton distribution functions and $\alpha_s(\mu_R)$ is the strong coupling at the renormalization  scale $\mu_R$. For details see Ref.~\cite{prompt1, jp1}. It is to be noted that in a complete and consistent NLO pQCD $[O (\alpha \alpha_s^2)]$ calculation important contribution to prompt photon result arises from various possible $ 2 \rightarrow 3 $ process  for the direct as well as for fragmentation processes~\cite{prompt1, vogel}. 

JETPHOX is a widely used parton level Monte Carlo event generator that provides prompt photon production cross section for high energy hadronic collisions. It allows the study of direct and fragmentation photon production at NLO accuracies using PDFs and fragmentation functions (FF).
One of the important features of JETPHOX framework is the implementation of the isolation criterion similar to experimental analysis and which can be varied to study its impact on the isolated photon spectra.
The isolation cone radius $R$ is defined as, 
\begin{equation}
R \ =  \ \sqrt {{\Delta \eta}^2  \ + \ {\Delta \phi}^2 \ }   \\
\end{equation}
The total transverse momentum of all particles within this isolation cone is then summed and required to be below a specified threshold ($p_{\rm{T}}^{\rm{cut}}$).
We have used the CT14  PDF set ~\cite{lhapdf} and the EPS09 parameterization of nuclear shadowing function~\cite{eps09} to estimate the prompt photon spectra. The BFG II ~\cite{frag} fragmentation function has been used to describe parton to photon fragmentation. The factorisation, renormalisation, and fragmentation scales were all set equal to the photon transverse momentum, i.e., $\mu_{f} = \mu_{R} = \mu_{F} = p_{\rm{T}}$.

The parameter $p_{\rm{T}}^{\rm{cut}}$ plays a central role in applying the isolation criterion for high $p_T$ prompt photon analyses. In experimental studies $p_{\rm{T}}^{\rm{cut}}$ is sometimes chosen as a fixed value, while in other cases it is defined as a fraction of the photon $p_T$. In this study,  we consider two commonly used isolation prescriptions, a fixed value of $p_{\rm{T}}^{\rm{cut}} = 2$ GeV and a dynamic choice of $p_{\rm{T}}^{\rm{cut}} = 0.1p_T$, both applied for a cone radius of $R=0.4$.

\section{Results and Discussions}
The transverse momentum spectra of prompt photons in $pp$ collisions at $\sqrt{s} = 7$ TeV calculated at NLO pQCD using the JETPHOX framework are shown in Fig.~\ref{fig1}. Results obtained with two isolation criteria, $p_{\rm{T}}^{\rm{cut}} = 2$ GeV and $p_{\rm{T}}^{\rm{cut}} =0.1 p_T$  are compared with ALICE measurements~\cite{alice_pp_7TeV}.  The data to JETPHOX ratio is shown in the lower panel of the figure. For each set of photon spectra five million events are generated within the central rapidity region $|\eta| < 0.27$. Both isolation conditions show reasonable agreement with the experimental data across a wide $p_T$ range. Furthermore, calculations without applying any isolation criterion are also shown for comparison.

The prompt photon spectra from Pb+Pb collisions using JETPHOX are also found to explain the CMS 5.02A TeV data ~\cite{cms_PbPb_5TeV} at the LHC reasonably well using the two different isolation criteria (see Fig.~\ref{fig2}) in the region $p_T > 25$ GeV and for $|\eta| < 1.44$. All calculations are shown for minimum bias events with the cross section expressed in per nucleon. The prompt spectrum without any isolation cut is also included in the figure. The JETPHOX framework has also been found to explain well the ALICE low $p_T$ ($< 15$ GeV) photon data from Pb+Pb collisions~\cite{alice2.76} at various centrality bins (see Fig.6 of Ref~\cite{fcc}).

\begin{figure}
\begin{center}
\includegraphics*[width=9.0cm,clip=true,angle=0]{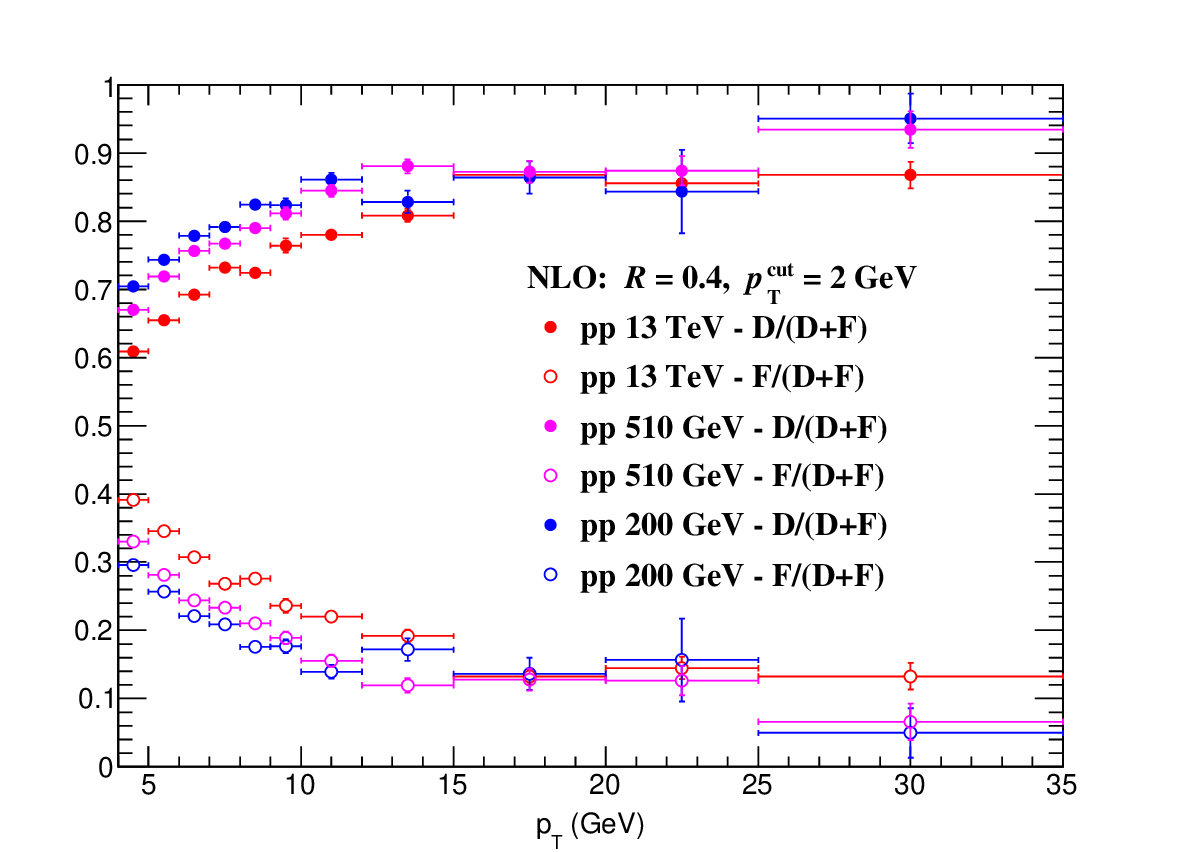}
\includegraphics*[width=9.0cm,clip=true,angle=0]{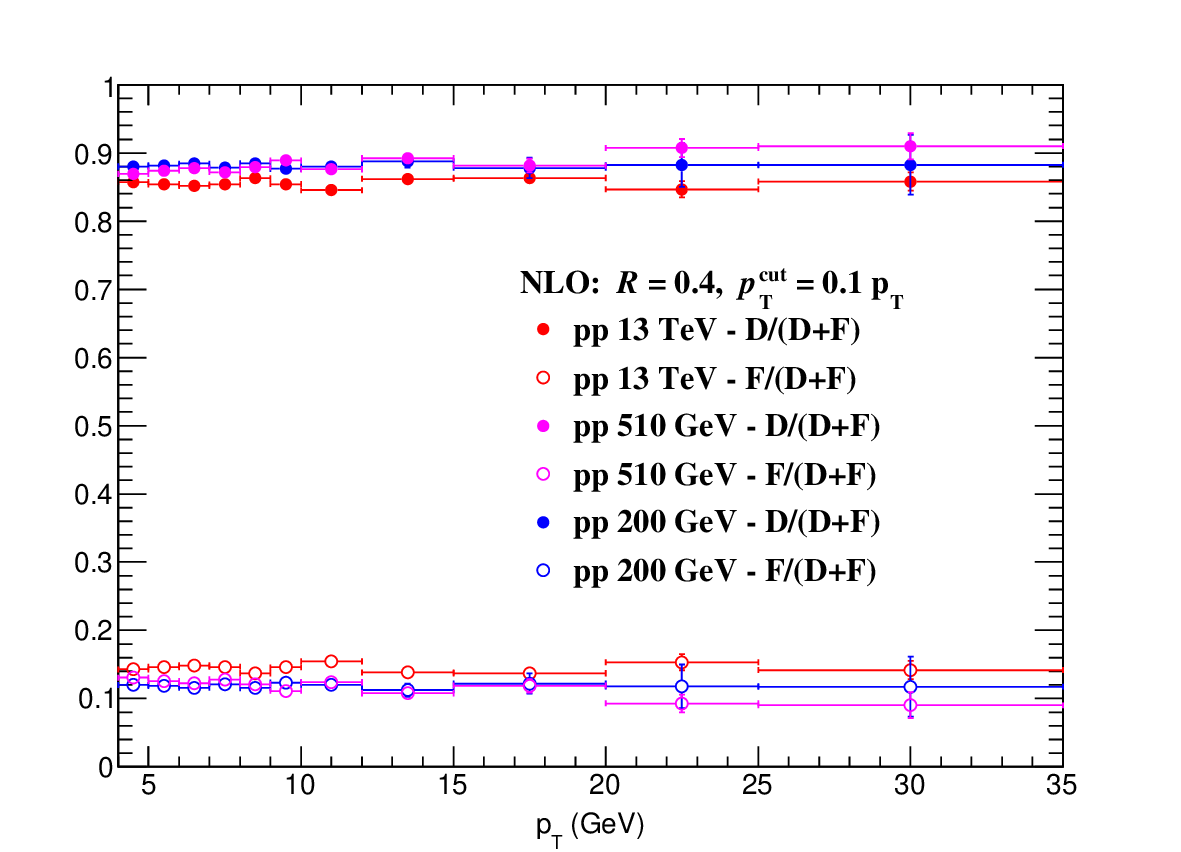}
\caption{Relative contributions of direct (D/(D+F)) and fragmentation (F/(D+F)) parts in total prompt production for isolation criterion [upper panel] $p_T^{\rm cut} = 2$ GeV and [lower panel] $p_T^{\rm cut} = 0.1 p_T$ for $pp$ collisions at various center of mass energies for $|\eta| < 0.5$ using JETPHOX~\cite{jp1}.} 
\label{fig3}
\end{center}
\end{figure}

 The JETPHOX calculations are found to be in good agreement with the PHENIX GeV $pp$ data for $p_T \ >$ 10 GeV [see Fig.~\ref{fig2.1}]. At lower transverse momenta ($p_T < \ $10 GeV), the theoretical results tend to underpredict the data as also reported earlier in Ref.~\cite{phenix_pp_510GeV}.

Fig.~\ref{fig3} shows the $p_T$ dependent relative contributions of the direct and fragmentation components to the total prompt photon production in $pp$ collisions. Results are presented for two different isolation criteria at center of mass energies of 200 GeV,  510 GeV and 13 TeV. The direct component dominates contributing about 60–90\% of the yield in the range 4 to 15  GeV $p_T$ value for the 13 TeV collisions and for $p_T^{\rm cut} = 2$ GeV. Beyond this $p_T$ range, the relative contribution from fragmentation photons becomes negligible. For 200 GeV collisions, the relative contribution of the direct component is found to be slightly larger than that at 13 TeV in the region $p_T < 15$ GeV. However, at both energies the fragmentation contribution becomes more prominent at lower $p_T$ region in the total spectra when the isolation cut is set to $p_T^{\rm cut} = 2$ GeV. The 510 GeV results are found to be consistent with those at 200 GeV as expected.


\begin{figure}
\begin{center}
\includegraphics*[width=8.8cm,clip=true,angle=0]{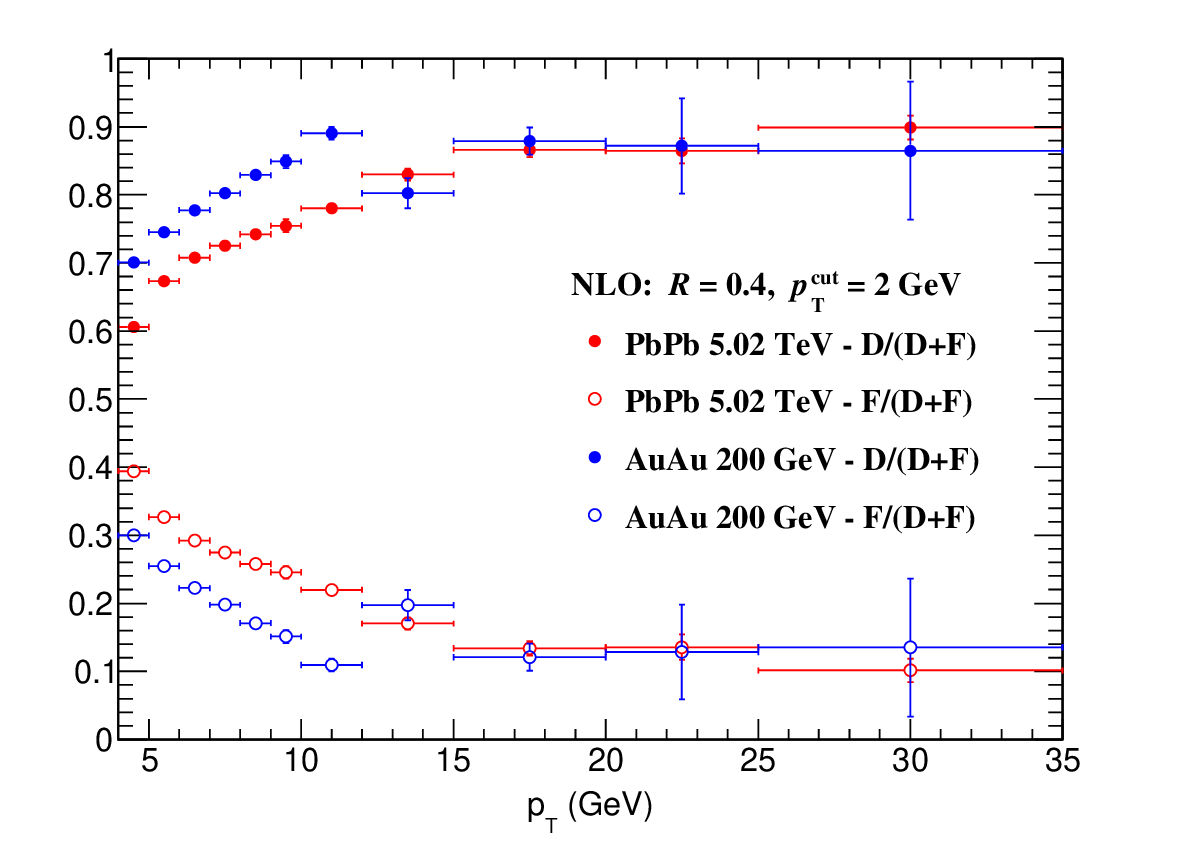}
\includegraphics*[width=8.8cm,clip=true,angle=0]{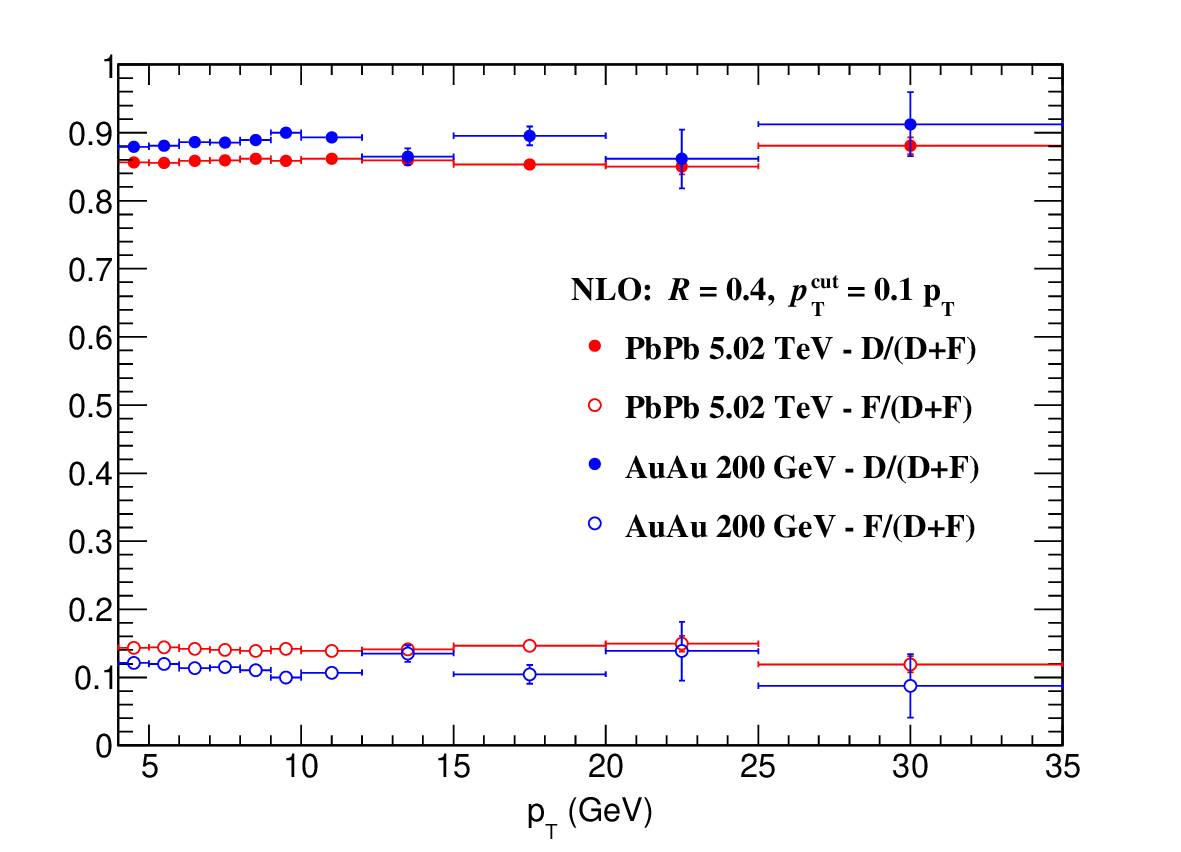}
\caption{ Relative contributions of direct (D/(D+F)) and fragmentation (F/(D+F)) parts in total prompt production for isolation criterion [upper panel] $p_T^{\rm cut} = 2 \ p_T$ and [lower panel] $p_T^{\rm cut} = 0.1 p_T$ for heavy ion collisions at different center of mass energies  for  $|\eta| < 0.5$ using JETPHOX~\cite{jp1}.} 
\label{fig4}
\end{center}
\end{figure}


\begin{figure}
\begin{center}
\includegraphics*[width=9.0cm,clip=true,angle=0]{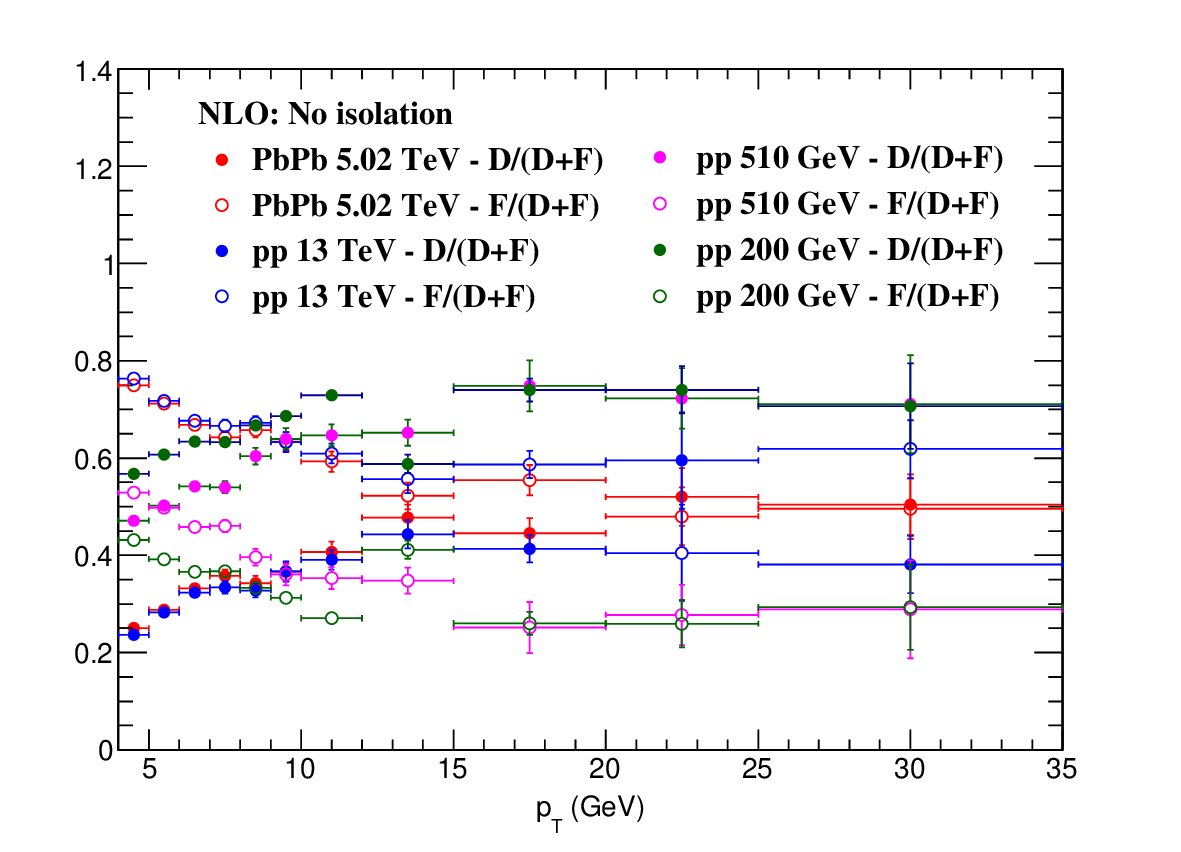}
\caption{Relative contributions of direct (D/(D+F)) and fragmentation (F/(D+F)) parts in total prompt production without any isolation cut at LHC and RHIC energies for $|\eta| < 0.5$ using JETPHOX~\cite{jp1}. The $Pb+Pb$ calculations are for the 0-100\% centrality class.}
\label{fig4.5}
\end{center}
\end{figure}

For $p_T^{\rm cut} = 0.1p_T$, the relative contributions of the direct and fragmentation components show no clear sensitivity to the photon $p_T$ or beam energy in the region $p_T > 4$ GeV. As illustrated in the lower panel of Fig.~\ref{fig3} the fragmentation contribution is strongly suppressed throughout the entire $p_T$ range.

The relative contributions of the direct and fragmentation components for heavy ion collisions are shown in Fig.~\ref{fig4} covering energies from 200A GeV Au+Au collisions at RHIC to 5.02 TeV Pb+Pb collisions at the LHC. The results from $AA$ collisions show similar qualitative nature to the results from $pp$ collisions. The relative contributions show beam energy dependence only for the $p_T^{\rm cut} = 2$ GeV isolation criterion.

These results show that the choice of the isolation criterion  affects the relative contributions of the direct and fragmentation components significantly. The fixed cut of $p_T^{\rm cut} = 2$ GeV shows a pronounced dependence on the beam energy and $p_T$, with higher beam energies resulting in a larger fraction of fragmentation photons contrary to the dynamic $p_T^{\rm cut}$. 

The prompt photon results without any isolation cut exhibit a qualitatively different $p_T$ dependent nature particularly in the region $p_T < 15$ GeV at LHC compared to the results obtained with isolation cuts. As shown in Fig.~\ref{fig4.5}, the fragmentation photons dominate the prompt photon spectrum when no isolation cut is applied. Consequently, the relative contribution from the fragmentation component is significantly larger than that of the direct component for $p_T < 10$ GeV while at higher $p_T$ both contributions become comparable as illustrated in the figure. For RHIC energy ($pp$ collisions at 200 and 510 GeV both) however, the direct components mostly dominate the spectra even without any isolation cut in the entire $p_T$ region shown in the figure.

\begin{figure}
\begin{center}
\includegraphics*[width=9.0cm,clip=true]{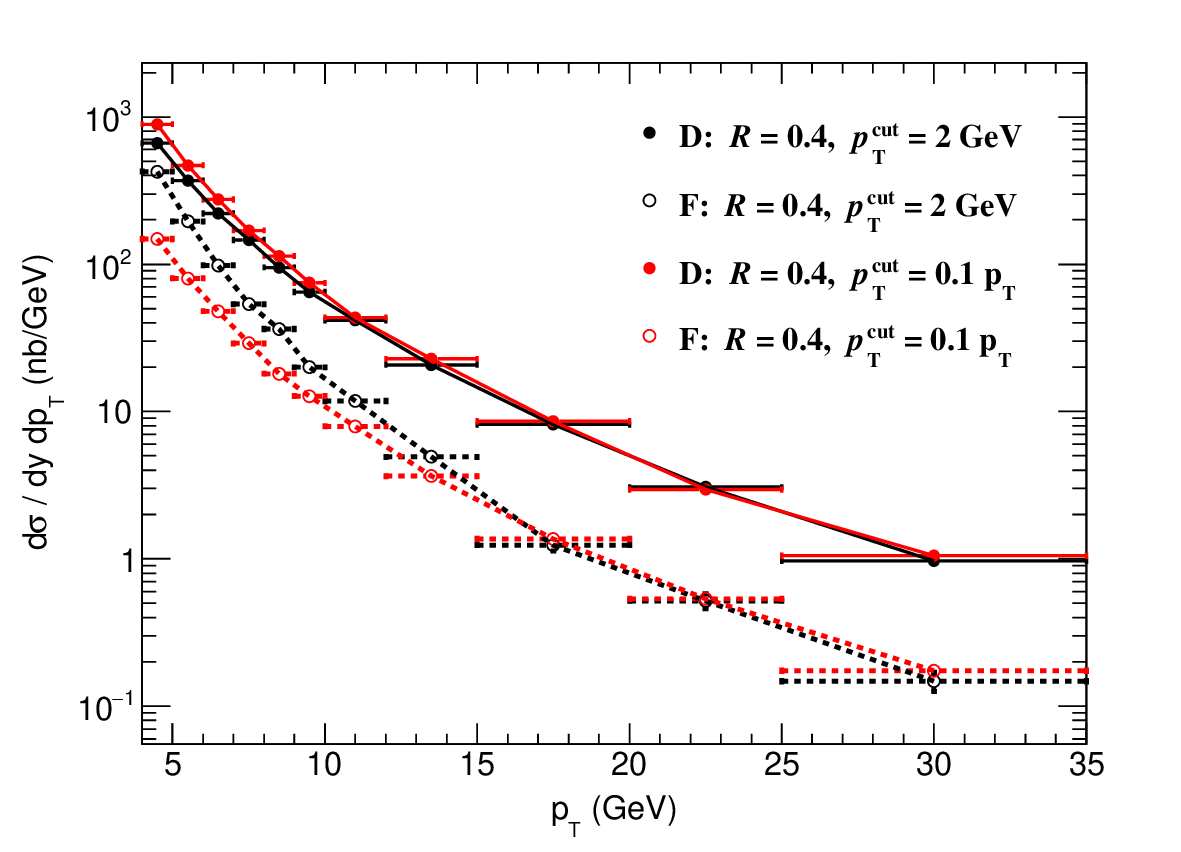}
\caption{Direct (D) and fragmentation (F) contributions to the total prompt photon production using two different isolation criteria for $pp$ collisions at 13 TeV at the LHC for $|\eta| < 0.5$.}
\label{fig5}
\end{center}
\end{figure}

\begin{figure}
\begin{center}
\includegraphics*[width=9.0cm,clip=true]{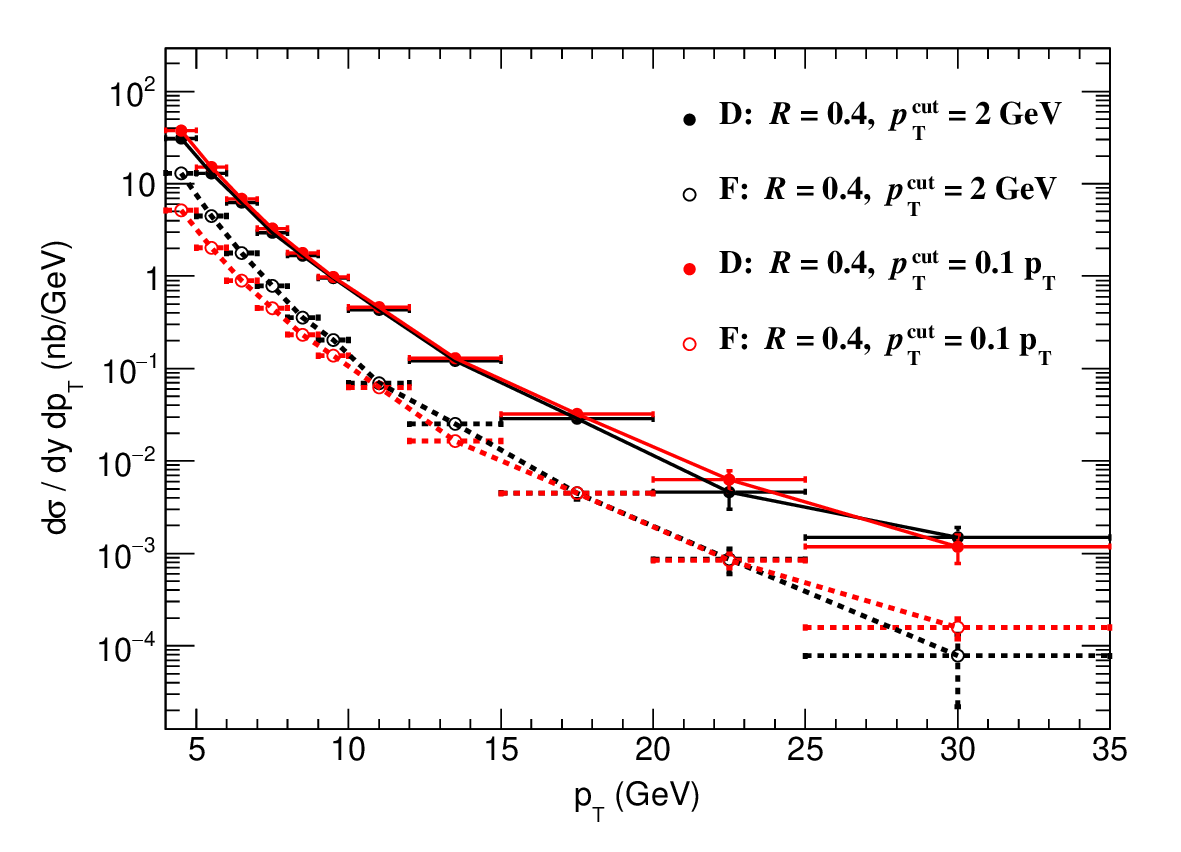}
\caption{Direct (D) and fragmentation (F) contributions to the total prompt photon production for using two different isolation criteria for $pp$ collisions at 200 GeV at RHIC for $|\eta| < 0.5$.}
\label{fig6}
\end{center}
\end{figure}

The cross sections of the direct and fragmentation components as functions of $p_T$ for $pp$ collisions at beam energies of 13 TeV and 200 GeV are shown in Figs.~\ref{fig5} and~\ref{fig6} respectively. These figures show results on a quantitative scale based on the two different isolation criteria.

At 13 TeV, the fragmentation contribution with the isolation criterion $p_T^{\rm cut} = 2$ GeV is about 2.5--3 three times larger than that with $p_T^{\rm cut} = 0.1p_T$ at $p_T = 4$ GeV. The difference between the two isolation criteria is pronounced in the  $p_T <$ 15 GeV region for fragmentation photons and it gradually decreases with increasing $p_T$.
The direct contribution also shows a slight variation between the two isolation criteria  which  affects the NLO $2 \rightarrow 3$ processes.
The fragmentation part from $pp$ collisions at 200 GeV shows a factor of 2--2.5 difference between the results from two different isolation cuts at 4 GeV $p_T$ value and the difference is found to be  significant in the range $p_T <$ 10 GeV.

\begin{figure}
\begin{center}
\includegraphics*[width=9.0cm,clip=true]{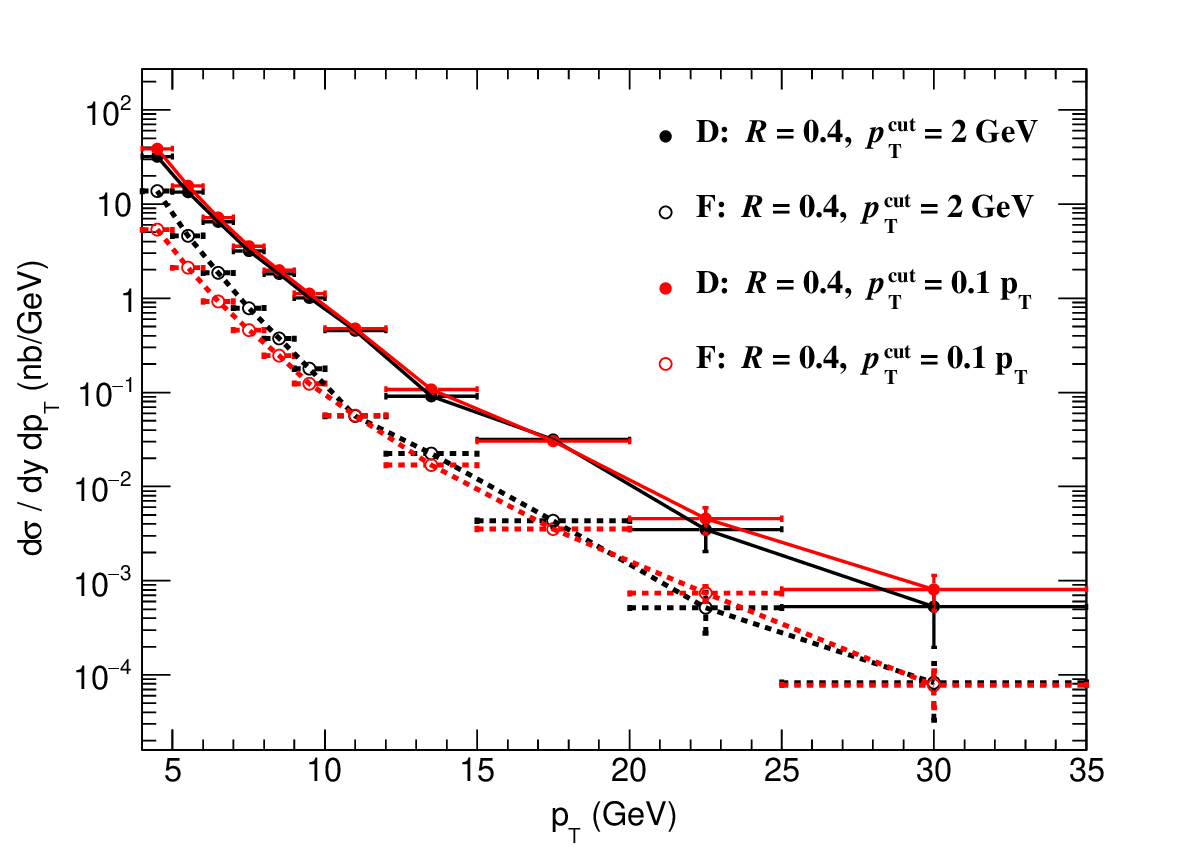}
\caption{Direct (D) and fragmentation (F) contributions to the total prompt photon production for using two different isolation criteria for Au+Au collisions at 200A GeV at RHIC for 0--100\% centrality class and $|\eta| < 0.5$.}
\label{fig7}
\end{center}
\end{figure}

\begin{figure}
\begin{center}
\includegraphics*[width=9.0cm,clip=true,angle=0]{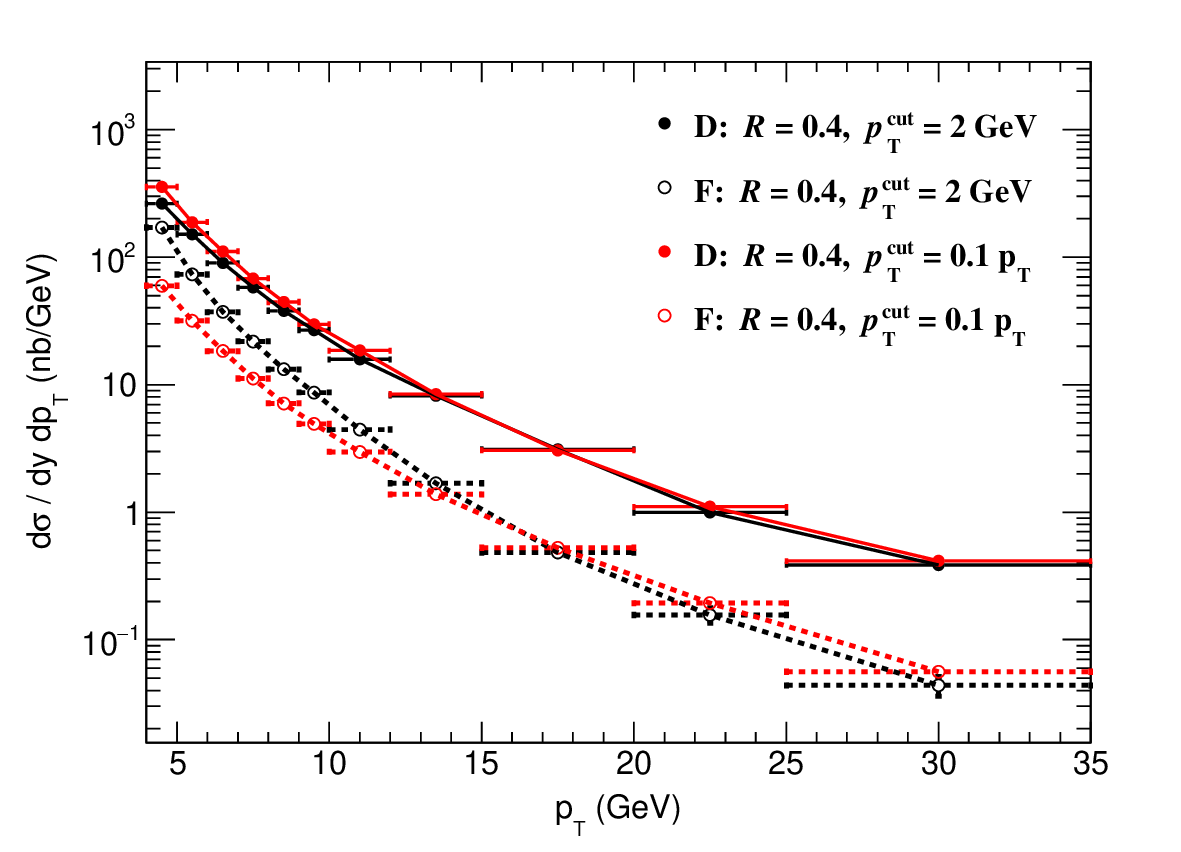}
\caption{Direct (D) and fragmentation (F) contributions to the total prompt photon production for using two different isolation criteria for Pb+Pb collisions at 5.02A TeV at the LHC for 0--100\% centrality class and $|\eta| < 0.5$.}
\label{fig8}
\end{center}
\end{figure}

For heavy ion collisions, the effect of the isolation cuts to the fragmentation component is found to be similar to that in $pp$ collisions as shown for 200A GeV Au+Au and 5.02A TeV Pb+Pb systems in Figs.~\ref{fig7} and~\ref{fig8}. The difference between the two isolation criteria amounts to roughly a factor of 3 for Pb+Pb collisions at 5.02A TeV and about a factor of 2.5 for Au+Au collisions at 200A GeV estimated at $p_T = 4$ GeV.

\begin{figure}
\begin{center}
\includegraphics*[width=9.0cm,clip=true,angle=0]{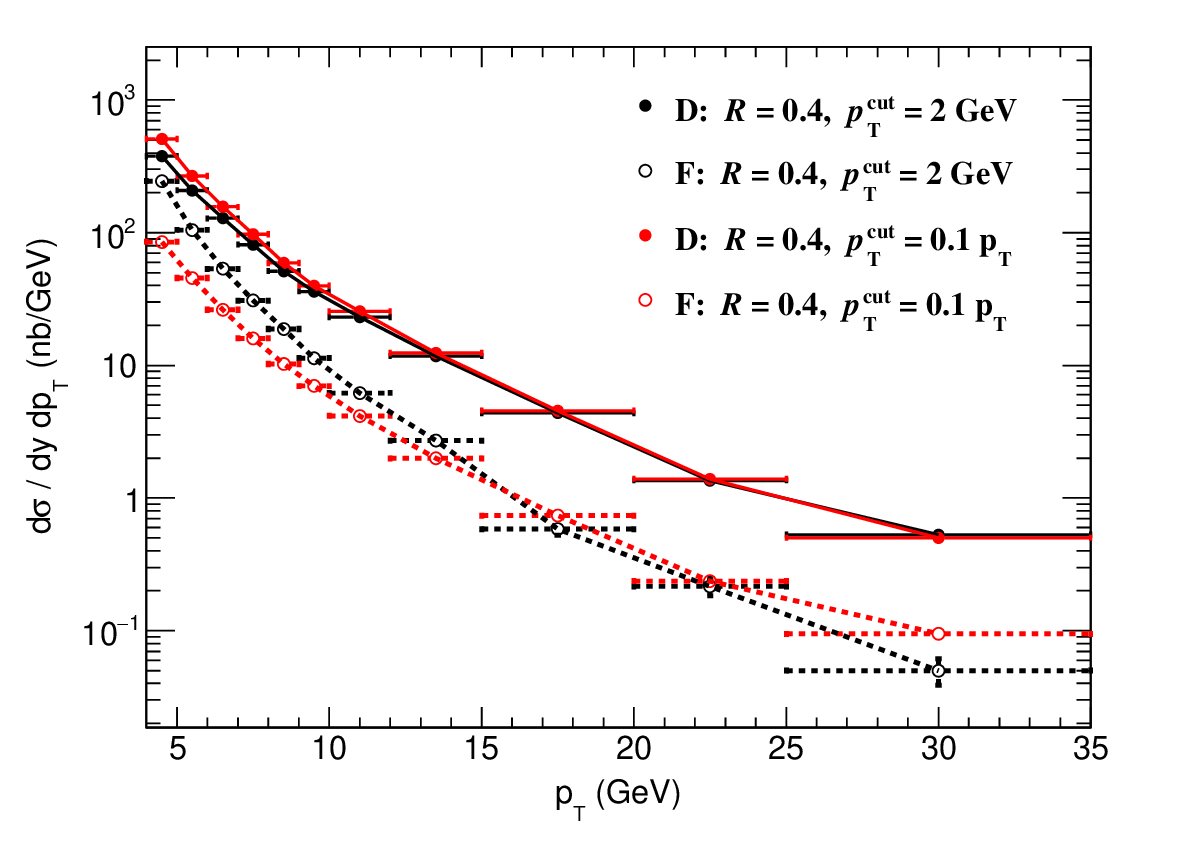}
\caption{Direct (D) and fragmentation (F) contributions to the total prompt photon production for using two different isolation criteria for O+O collisions at 7A TeV at the LHC. The calculations are shown for 0-100\% centrality class and $|\eta| < 0.5$.}
\label{fig9}
\end{center}
\end{figure}

\begin{figure}
\begin{center}
\includegraphics*[width=9.0cm,clip=true,angle=0]{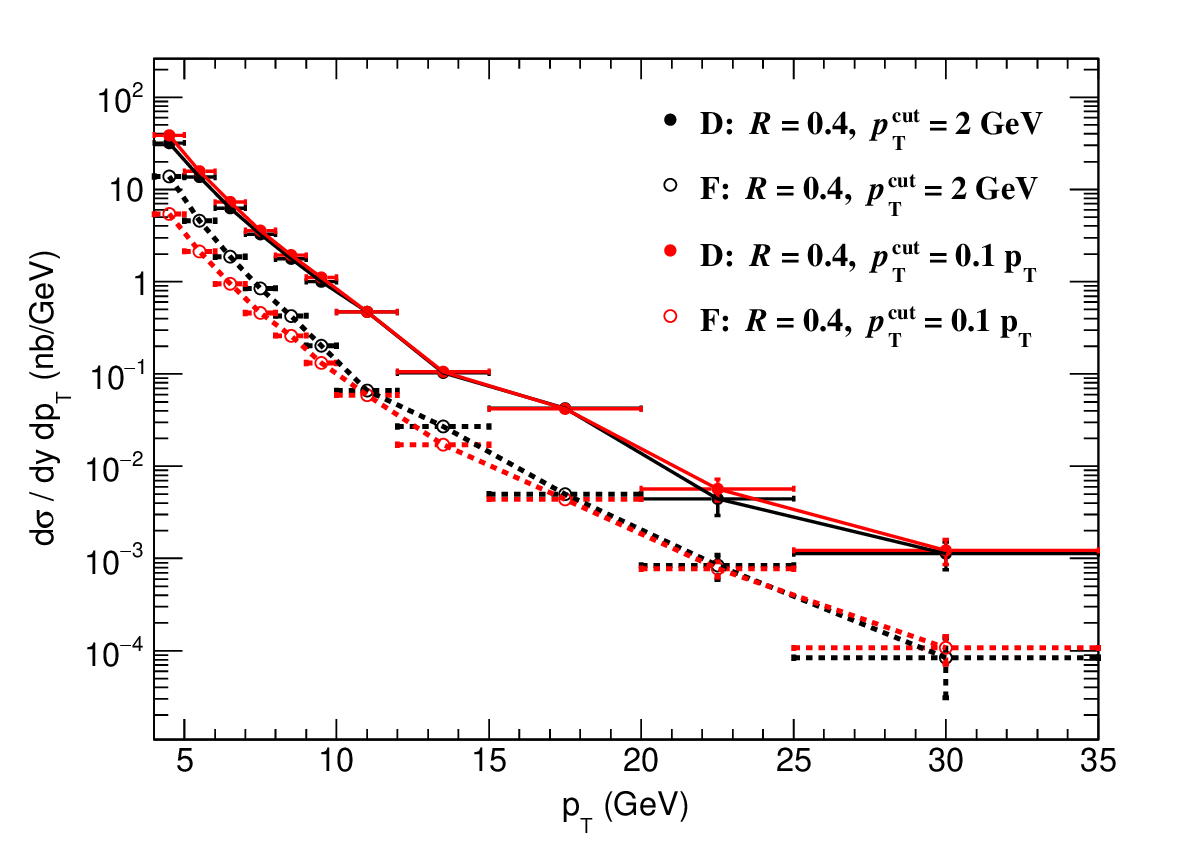}
\caption{Direct (D) and fragmentation (F) contributions to the total prompt photon production for using two different isolation criteria for Ru+Ru collisions at 200A GeV at the RHIC. The calculations are shown for 0-100\% centrality class and $|\eta| < 0.5$.}
\label{fig10}
\end{center}
\end{figure}

Collisions of oxygen nuclei at top LHC energies and isobaric collisions at RHIC are expected to provide valuable and complementary insights to those obtained earlier from heavy ion systems such as Au+Au at RHIC and Pb+Pb at the LHC~\cite{oo, oo1, isobar, isobar1}. The prompt photon spectra from $^{16}$O+$^{16}$O collisions at $\sqrt{s_{NN}}=7$ TeV exhibit a qualitatively similar behavior to Pb+Pb collisions at the similar center of mass energy [see Fig.\ref{fig9}]. Additionally, results from $^{96}$Ru+$^{96}$Ru collisions at 200A GeV show similar trends to those from Au+Au collisions at the same energy [Fig.\ref{fig10}] (EPS08 nPDF set is used for this particular case~\cite{eps08}). 

These observations further demonstrate that the influence of the nuclear medium is marginal compared to the choice of isolation criterion in the analysis of prompt photons as the results at a given center of mass energy remain similar for a particular isolation cut. 

\begin{figure}
\includegraphics[width=90mm,trim={0 0 0 0},clip]{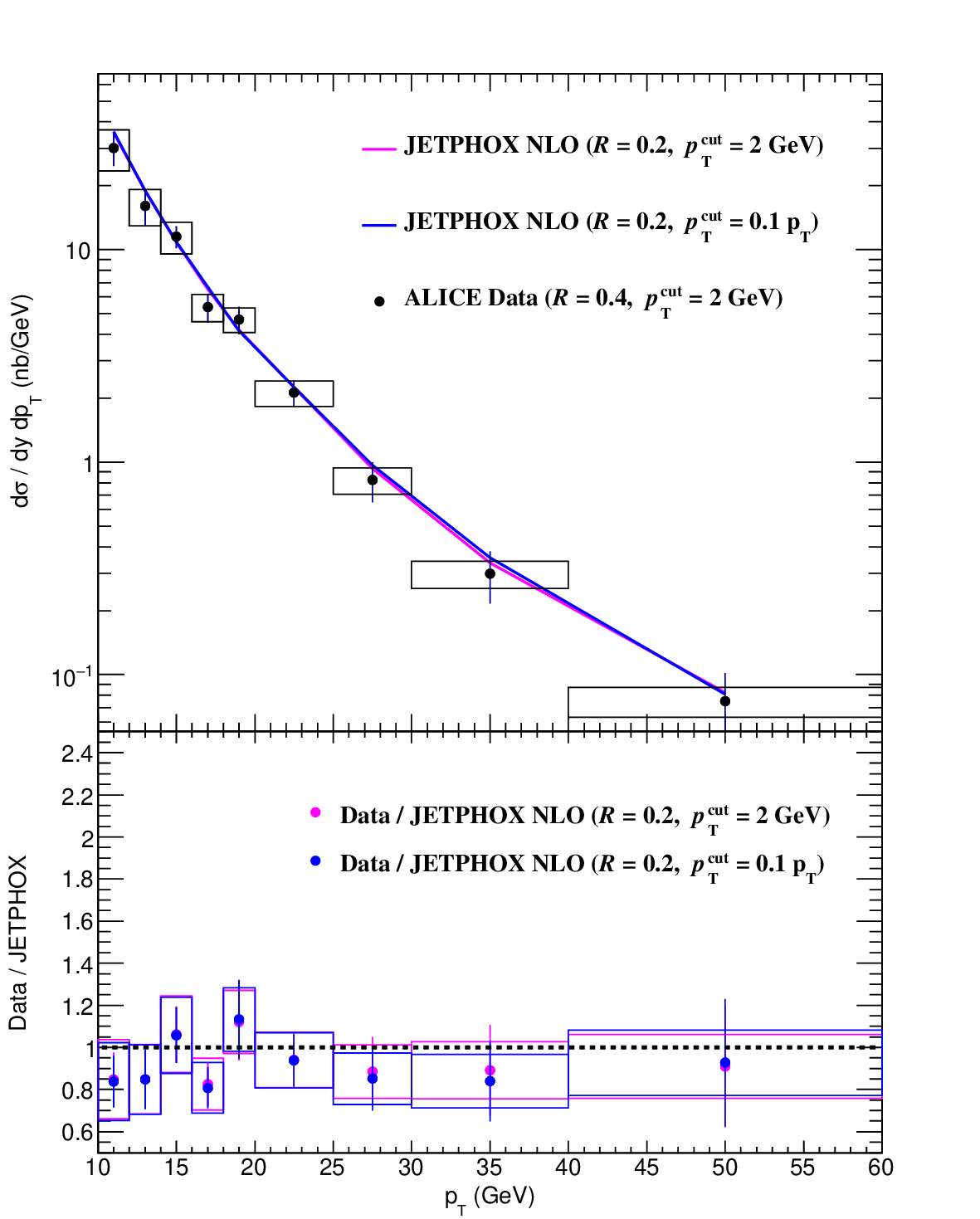}  
\caption{\label{fig1a} Prompt photon cross section in $pp$ collisions at $\sqrt{s}=7$ TeV for two different isolation criteria using $R=0.2$ along with ALICE data~\cite{alice_pp_7TeV} in the rapidity range $|\eta|<0.27$.}
\end{figure}

The isolation cone radius is also varied in some experimental analyses, typically within the range $R =$ 0.2--0.4~\cite{atlas_7tev, atlas_13_1, atlas_13_2, atlas_pp_13, alice_PbPb_5.02, cms13}. In Fig.~\ref{fig1a} we compare our results with the ALICE data using an isolation cone radius of $R =$ 0.2 for two different isolation criteria in 7 TeV $pp$ collisions at the LHC. The corresponding JETPHOX results are found to be consistent with the prompt photon results within uncertainties as shown in the figure.

\begin{figure}
\begin{center}
\includegraphics*[width=9.0cm,clip=true]{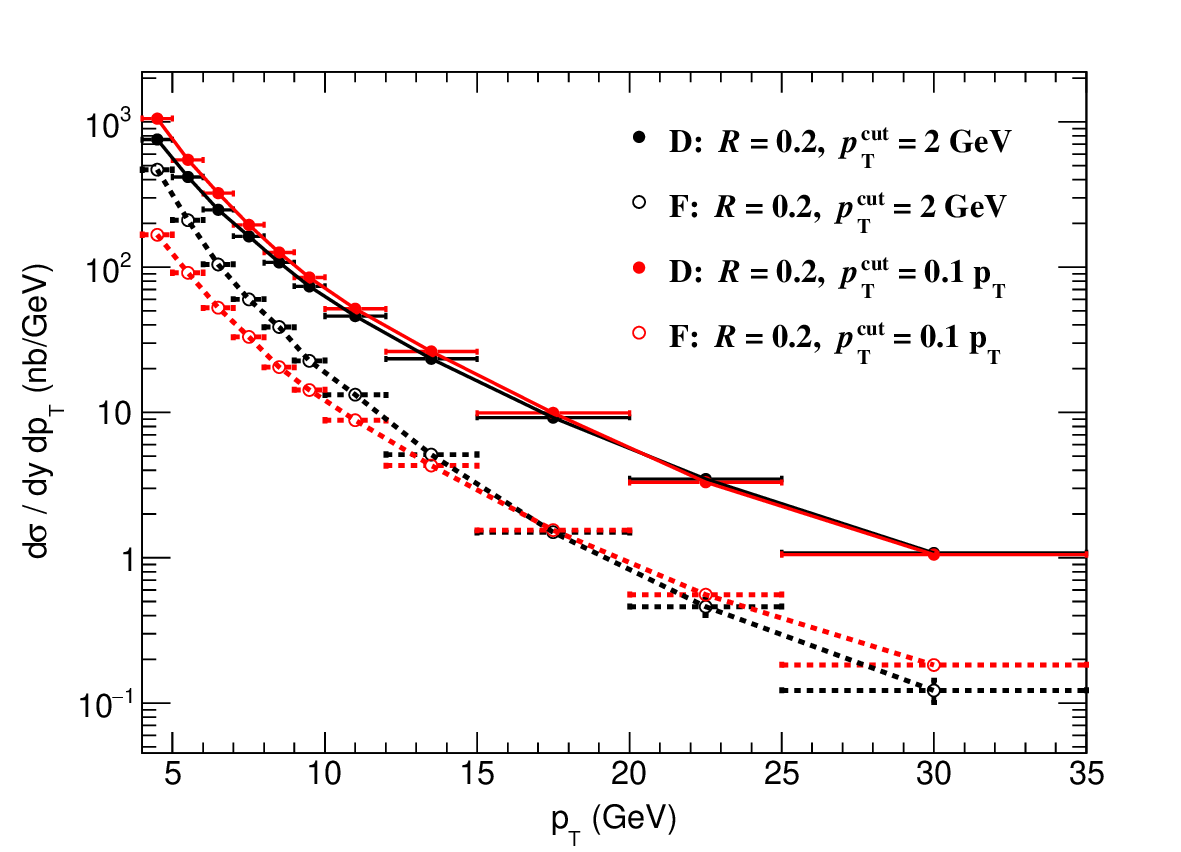}
\caption{Direct (D) and fragmentation (F) contributions to the total prompt photon production using two different isolation criteria using $R =$ 0.2 for $pp$ collisions at 13 TeV at the LHC in the rapidity range $|\eta| < 0.5$.}
\label{fig5a}
\end{center}
\end{figure}

\begin{figure}
\begin{center}
\includegraphics*[width=9.0cm,clip=true]{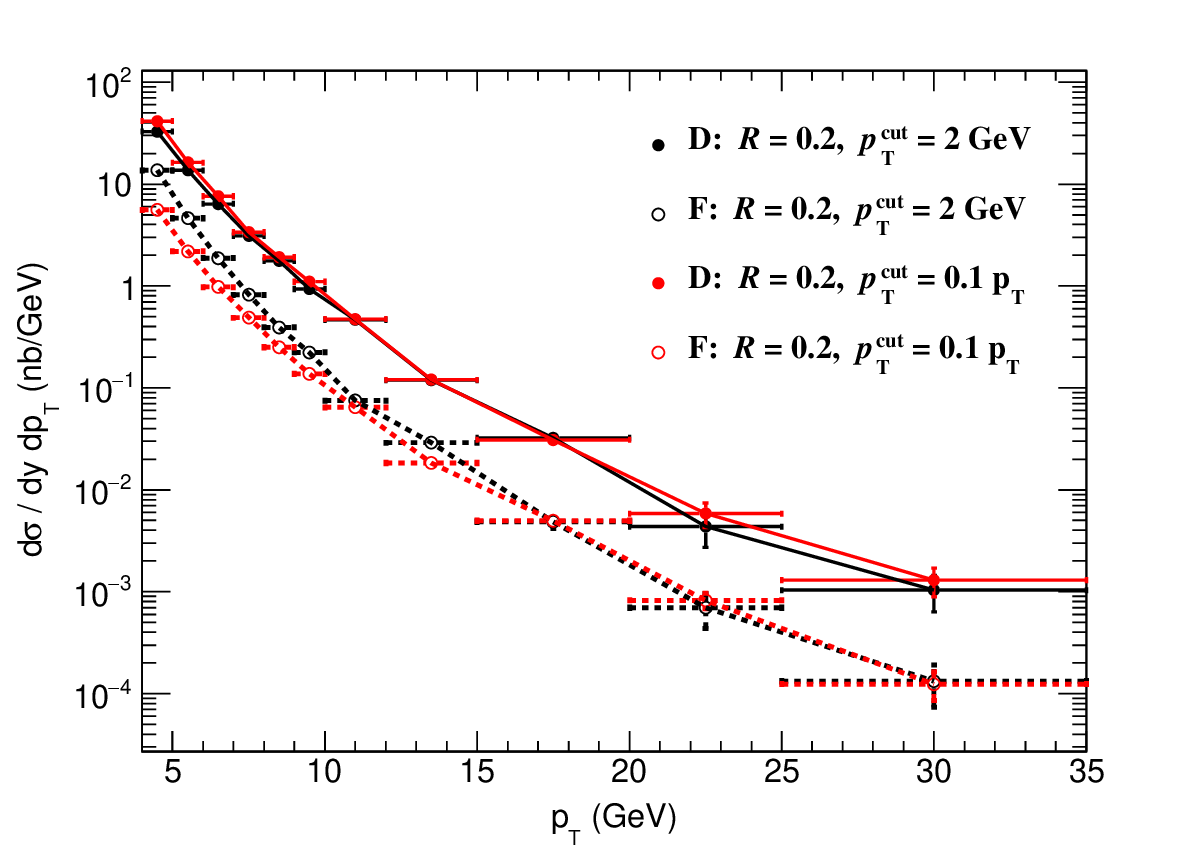}
\caption{Direct (D) and fragmentation (F) contributions to the total prompt photon production for using two different isolation criteria using $R =$ 0.2 for $pp$ collisions at 200 GeV at RHIC in the rapidity range $|\eta| < 0.5$.}
\label{fig6a}
\end{center}
\end{figure}

The direct and fragmentation contributions for $R =$ 0.2 are also presented in Figs.~\ref{fig5a} and~\ref{fig6a} for $pp$ collisions at 13 TeV and 200 GeV respectively, under the two isolation conditions. These results are found to be similar to those shown earlier in Figs.~\ref{fig5} and~\ref{fig6} for  $R =$ 0.4.

It can be concluded from these results that the prompt photons are more sensitive to the $p_T^{\rm cut}$ than to the choice of isolation cone radius $R$ (in the range 0.2--0.4) particularly in the fragmentation contribution for $p_T$ region below 15 GeV. 

\section{summary and conclusions}
The measurement of prompt photons in proton-proton collisions provides a crucial baseline for heavy ion measurements, thereby helping to isolate the medium induced effects. A precise understanding of the different photon sources and their relative contributions are therefore crucial for an accurate interpretation of the results. 

Prompt photon production has been  studied in detail at various center of mass energies and collision systems at both RHIC and the LHC in this work. The direct and fragmentation components of the photon spectrum are investigated using the JETPHOX framework which successfully describes the experimental data at RHIC and LHC up to very large values of $p_T$.

Two different isolation criteria typically used in high energy experiments;  $p_T^{\rm cut} = \ 2 $ GeV and $p_T^{\rm cut} = 0.1p_T$, within a cone of radius $R=0.4$ are used to study the relative contributions of the direct and fragmentation parts of prompt photon production.
The isolation criteria is found to play a crucial role in the prompt photon analysis especially in the $p_T$ region 4--15 GeV and it also depends on the center of mass energy of the collisions.

The $p_T$ dependent isolation criterion $p_T^{\rm cut} = 0.1p_T$ with $R=0.4$  is found to be more effective in suppressing the fragmentation contribution compared to the fixed cut of $p_T^{\rm cut} = 2$ GeV in the  $p_T$ range 4--15 GeV. 
Varying the isolation cone radius from 0.4 to 0.2 shows that the results are more sensitive to the $p_T^{\rm cut}$ than to the cone radius $R$ below 15 GeV for the fragmentation contribution.

Therefore, a judicious choice of isolation criterion is crucial for accurately analyzing isolated photon spectra in this region. 
At higher transverse momenta ($p_T > $15 GeV) both isolation cut effectively suppresses the fragmentation contribution. However, below this $p_T$ region the degree of suppression is sensitive to the used $p_T^{\rm cut}$ and is also found to be sensitive to the photon $p_T$ and the center of mass energy.

It is to be noted that jet–photon conversion can make a significant contribution to the photon spectra in Au+Au collisions at 200A GeV particularly for $p_T <\ $ 8 GeV~\cite{jet_phot}. Consequently, a precise estimate of the jet–photon conversion contribution is essential when analyzing the photon data below that $p_T$ range and comparing them with pQCD-based model calculations.

Prompt fragmentation photons  are expected to produce a small but positive elliptic flow in the low and intermediate $p_T$ regions along with the thermal contribution~\cite{frag_v2}. Whereas, the direct component of the prompt contribution does not contribute to anisotropic flow. Consequently the photon elliptic flow parameter can be influenced by variations in the fragmentation contribution arising from different isolation criteria.  Additionally, the nuclear modification factor $R_{\rm AA}$ for photons is also expected to be sensitive to the choice of isolation cut.

In conclusion, these results demonstrate that the choice of isolation criterion is crucial for a precise interpretation of prompt photon production and related observables in relativistic nuclear collisions.

\begin{acknowledgments}
The authors sincerely appreciate the insightful correspondences with Jean-Philippe Guillet concerning JETPHOX. We gratefully acknowledge Mr. Abhishek Seal for his assistance and support in running the JETPHOX code during this study. SC further acknowledges the use of the GRID computing facilities at VECC.
\end{acknowledgments}

\end{document}